\documentclass[openacc]{rstransa}
\usepackage{amsmath}

\begin{document}

\title{Study on Green's function on TI surface}

\author{
Bo Lu, Yukio Tanaka}

\address{Department of Applied Physics, Nagoya University, Nagoya 464-8603, Japan\\
and CREST, Japan Science and Technology Corporation (JST), Nagoya 464-8603, Japan }

\subject{Topological Insulator, Green's Function, Josephson effect}


\corres{Bo Lu\\
\email{dr.lv.bo@gmail.com}}

\begin{abstract}
In theory of superconducting junctions, Green's function
has an important role
to obtain Andreev bound states, local density of states and
Josephson current in a systematic way.
In this article, we show how to construct Green's function on the surface
of topological insulator following McMillan's formalism where
the energy spectrum of electrons obeys a linear dispersion.
For a model of superconductor (S)/ferromagnet (F)/normal metal (N) junction,
we show that the generation of Majorana Fermion gives rise to
the enhanced local density of states and pair amplitude of
odd-frequency pairing.
We also derive an extended Furusaki-Tsukada's formula of d.c. Josephson current in S/F/S junctions.
The obtained Josephson current depends on the direction and magnitude of
the magnetization.

\end{abstract}


\begin{fmtext}








\end{fmtext}

\maketitle

\section{Introduction}

About 50 years before, McMillan has developed a theory of Green's function
\cite{McMillan} to study proximity effect in normal metal / superconductor
junctions. McMillan's theory of Green's function has been known as a
classical and standard one to study proximity or Josephson effect in
superconducting junctions. This theory is available in a ballistic regime
where the charge carriers, electrons or holes, can be described by coherent
wave functions, known as Bogoliubov quasiparticles. The Green's function is
obtained by composing these wave functions in a systematic way. Thus, it is
a very basic way to obtain Green's function for ballistic junction
problems.

Although there exist other approaches to solve Green's function such as
recursive technique, McMillan's formalism has its advantages and can make
simplifications in certain problems, such as Josephson effect. It is also
known that the surface Green's function can also be found by McMillan's method
\cite{Datta98} which is used as an initial condition in the recursive
technique\cite{Furu98}. In its application of d.c. Josephson current, this
method and the derived formula provide a clear physical correspondence
between quasiparticle transport and supercurrent flow. In this article, we
briefly review its mathematical framework and present a simple example to
demonstrate how it works in calculating proximity effect and Josephson
current in superconducting heterostructures on the surface of topological
insulator (TI).

Before we start discussion, let us survey a history of this theory and its
applications to superconducting junctions. The original work by McMillan\cite%
{McMillan} is to study the proximity effect of normal metal / conventional BCS superconductor
junctions. After Blonder, Tinkham and Klapwijk's theory\cite{BTK}, it becomes
popular to study the McMillan's Green's function based on BdG Hamiltonian%
\cite{BdG}. In 1991, Furusaki and Tsukada\cite{FT} derived a formula to
calculate d.c.Josephoson current. The extension into $d$-wave junctions have been done in the late 90s \cite{Tanaka96,TanakaD97,Tanaka2000,Barash}.
Also theory of unconventional Josephson junctions available for
$p$-wave \cite{Asano00,Asano06,Brydon08} case have been developed.
Almost 10 years ago, it has been found that the McMillan's formalism is also
applicable in superconductors with linear dispersion like Graphene\cite%
{Asano08,Burst10}. Very recently, it has been
revealed that McMillan's formalism
is powerful to study the transport phenomena in new materials such as 1D
helical edge states \cite{Pablo15} or 2D surface states \cite{Benj,bYang} on
TI surface.

The rest of this paper is organized as follows. In section II, we will give
an intuitive introduction of the approach of McMillan's theory. In section III,
we present an example of constructing Green's function in an $s$-wave superconductor (S)/ferromagnet (F)/normal metal junction
on TI surface. In Section IV, we will study the S/F/S Josephson junction on TI surface. A
concluding remark is given in Section V.

\section{McMillan's Green's function}

We start by reviewing some general aspects of McMillan's formalism. The
retarded and advanced Green's functions are defined by
\begin{eqnarray}
G^{R}\left( \mathbf{x},\mathbf{x}^{\prime },t,t^{\prime }\right)
&=&-i\theta \left( t-t^{\prime }\right) \left\langle \left\{ \hat{\Psi}%
\left( \mathbf{x}\right) ,\hat{\Psi}^{\dag }\left( \mathbf{x}^{\prime
}\right) \right\} \right\rangle , \\
G^{A}\left( \mathbf{x},\mathbf{x}^{\prime },t,t^{\prime }\right)  &=&i\theta
\left( t^{\prime }-t\right) \left\langle \left\{ \hat{\Psi}\left( \mathbf{x}%
\right) ,\hat{\Psi}^{\dag }\left( \mathbf{x}^{\prime }\right) \right\}
\right\rangle .
\end{eqnarray}%
They follow equations%
\begin{equation}
\left[ E-\hat{H}\left( \mathbf{x}\right) \right] G^{R\left( A\right) }\left(
\mathbf{x},\mathbf{x}^{\prime },E\right) =\delta \left( \mathbf{x-x}^{\prime
}\right) ,  \label{left}
\end{equation}%
\begin{equation}
G^{R\left( A\right) }\left( \mathbf{x},\mathbf{x}^{\prime },E\right) \left[
E-\hat{H}\left( \mathbf{x}^{\prime }\right) \right] =\delta \left( \mathbf{%
x-x}^{\prime }\right) .  \label{right}
\end{equation}%
The operator $\mathcal{L}=E-\hat{H}$\textbf{\ }acts on the Green's function
from the left side in Eq.(\ref{left}), while it does from the right side in
Eq.(\ref{right}). Thus, $G^{R\left( A\right) }\left( \mathbf{x},\mathbf{x}%
^{\prime },E\right) $ is generally composed by $\bar{\Psi}\left( \mathbf{x}%
\right) =\sum\nolimits_{n}c_{n}\Psi _{n}\left( \mathbf{x}\right) $ and $\bar{%
\Phi}\left( \mathbf{x}^{\prime }\right) =\sum\nolimits_{m}d_{m}\Phi
_{m}\left( \mathbf{x}^{\prime }\right) $ with
\begin{equation}
G=\bar{\Psi}\left( \mathbf{x}\right) \bar{\Phi}^{T}\left( \mathbf{x}^{\prime
}\right) .
\end{equation}%
$\Psi _{n}\left( \mathbf{x}\right) $ and $\Phi _{m}\left( \mathbf{x}^{\prime
}\right) $ satisfy
\begin{equation}
\mathcal{L}\left( \mathbf{x}\right) \Psi _{n}\left( \mathbf{x}\right) =0,
\label{element}
\end{equation}%
\begin{equation}
\Phi _{m}^{T}\left( \mathbf{x}^{\prime }\right) \mathcal{L}\left( \mathbf{x}%
^{\prime }\right) =0,  \label{trans}
\end{equation}%
where the index $m$ ($n$) is the quantum number and $c_{n}$ and $d_{m}$ are
coefficients which will be determined in the framework of McMillan's theory.
We can take the transpose on both sides of Eq.(\ref{trans})
\begin{equation}
\mathcal{L}\left( \mathbf{x}^{\prime }\right) ^{T}\Phi _{m}\left( \mathbf{x}%
^{\prime }\right) =0,  \label{conjuta}
\end{equation}%
To find the proper form of $\mathcal{L}\left( \mathbf{x}^{\prime }\right)
^{T}$, we should use the fact $\hat{k}_{x^{\prime },y^{\prime },z^{\prime
}}^{T}=-\hat{k}_{x^{\prime },y^{\prime },z^{\prime }}$, where $\hat{k}%
_{x,y,z}$ is a real space operator $\hat{k}_{x^{\prime },y^{\prime
},z^{\prime }}=-i\partial _{x^{\prime },y^{\prime },z^{\prime }}$. For
example, if we have 2 by 2 Hamiltonian
\begin{equation}
\hat{H}\left( \mathbf{x}^{\prime }\right) =\sum_{i}h_{i}\left( \hat{k}%
_{x^{\prime }},\hat{k}_{y^{\prime }},\hat{k}_{z^{\prime }}\right) \hat{\sigma%
}_{i},
\end{equation}%
where $\hat{\sigma}_{i=0,1,2,3}$ is Pauli matrix, $\hat{H}^{T}\left( \mathbf{%
x}^{\prime }\right) $ is then expressed by
\begin{equation}
\hat{H}^{T}\left( \mathbf{x}^{\prime }\right) =\sum_{i}h_{i}\left( -\hat{k}%
_{x^{\prime }},-\hat{k}_{y^{\prime }},-\hat{k}_{z^{\prime }}\right) \hat{%
\sigma}_{i}^{\ast }.  \label{hamilc}
\end{equation}%
For convenience, we define a "new" operator $\tilde{H}$ as
\begin{equation}
\tilde{H}\left( \mathbf{x}^{\prime }\right) =\hat{H}^{T}\left( \mathbf{x}%
^{\prime }\right) ,
\end{equation}%
As seen  from the 2 by 2 example of Eq.(\ref{hamilc}), one can find the
general relation that the wave function $\Phi $ of $\tilde{H}\left( \mathbf{x%
}\right) $ and the wave function $\Psi $ of $\hat{H}^{T}\left( \mathbf{x}%
\right) $ have the one-to-one correspondence%
\begin{equation}
\Phi _{k_{x},k_{y},k_{z}}^{T}\left( E\right) =\Psi
_{-k_{x},-k_{y},-k_{z}}^{\dag }\left( E\right) ,
\end{equation}%
or
\begin{equation}
\Phi _{k_{x},k_{y},k_{z}}\left( E\right) =\Psi _{-k_{x},-k_{y},-k_{z}}^{\ast
}\left( E\right) ,
\end{equation}%
where $k_{x},k_{y},k_{z}$ are quantum numbers. And we know that the Hilbert
spaces spanned by the eigenfunctions of $\tilde{H}$ and $\hat{H}$ have
one-to-one correspondence. Then, we have%
\begin{equation}
\left[ E-\hat{H}\left( \mathbf{x}\right) \right] G^{R\left( A\right) }\left(
\mathbf{x},\mathbf{x}^{\prime },E\right) =\delta \left( \mathbf{x-x}^{\prime
}\right) ,  \label{pro1}
\end{equation}%
\begin{equation}
\left[ E-\tilde{H}\left( \mathbf{x}^{\prime }\right) \right] \left[
G^{R\left( A\right) }\left( \mathbf{x},\mathbf{x}^{\prime },E\right) \right]
^{T}=\delta \left( \mathbf{x-x}^{\prime }\right) .  \label{pro2}
\end{equation}%
The Green's function of Eq.(\ref{pro1}) corresponds to the elementary
process Eq.(\ref{element}) and the Green's function of Eq.(\ref{pro2})
corresponds to the conjugate process Eq.(\ref{conjuta}). If the system has a
translational invariance along $y$- and $z$-directions, $G^{R}\left( \mathbf{%
x},\mathbf{x}^{\prime },E\right) =G_{k_{y},k_{z}}^{R}\left( x,x^{\prime
},E\right) e^{ik_{y}\left( y-y^{\prime }\right) +ik_{z}\left( z-z^{\prime
}\right) }$ is satisfied. Therefore, we must choose the solutions of Eq.(\ref%
{element}) with quantum number $\left( k_{y},k_{z}\right) $ and those of Eq.(%
\ref{conjuta}) with $\left( -k_{y},-k_{z}\right) $.

The specific components of $\bar{\Psi}\left( \mathbf{x}\right) $ $\left(
\bar{\Phi}^{T}\left( \mathbf{x}^{\prime }\right) \right) $ are determined by
the type of Green's function we seek. If we consider the retarded Green's
function, we need outgoing waves. In this case, Eq.(\ref{pro1}) means that
the impulse at $\mathbf{x}^{\prime }$ will propagate and affect the field at
$\mathbf{x}$. On the other hand, Eq.(\ref{pro2}) means that the impulse at $%
\mathbf{x}$ will propagate and affect the field at $\mathbf{x}^{\prime }$.
If we consider the advanced Green's function, we need incoming waves.

\section{Green's Function in S/FI/N junction on TI surface}

In this section, we show an example to construct McMillan's Green's
function. Let us consider an planer $s$-wave superconductor (S)/ferromagnet
(F)/normal metal (N) junction on the surface of TI. The Dirac-type
Hamiltonian is given by
\begin{equation}
\hat{H}=\left[
\begin{array}{cc}
h(k_{x},k_{y})+M & i\hat{\sigma}_{y}\Delta \\
-i\hat{\sigma}_{y}\Delta & -h^{\ast }(-k_{x},-k_{y})-M^{\ast }%
\end{array}%
\right] ,
\end{equation}%
where $h(k_{x},k_{y})=v_{f}(k_{y}\hat{\sigma}_{x}-k_{x}\hat{\sigma}_{y})-\mu
$. $\hat{\sigma}_{i=x,y,z}$ is the Pauli matrix in the spin space and $\mu $
is the chemical potential. The pair potential $\Delta $ is given by $\Delta
_{0}\Theta (-x)$ and the Hamiltonian has a translational invariance in the $%
y $-direction. Throughout this paper, we set $\hbar =1$. The exchange field
in F region is $M=\sum_{i=x,y,z}m_{i}\hat{\sigma}_{i}\Theta \left( x\right)
\Theta \left( L-x\right) $. For an elementary process, we solve four
eigenvectors $\hat{A}_{1}e^{ik^{+}x+ik_{y}y}$, $\hat{A}%
_{2}e^{-ik^{-}x+ik_{y}y}$, $\hat{A}_{3}e^{-ik^{+}x+ik_{y}y}$ and $\hat{A}%
_{4}e^{ik^{-}x+ik_{y}y}$ with%
\begin{equation}
\hat{A}_{1}=\frac{1}{Z_{1}^{\frac{1}{2}}}\left[
\begin{array}{c}
i \\
e^{i\theta _{+}} \\
-e^{i\theta _{+}}\gamma \\
i\gamma%
\end{array}%
\right] ;\hat{A}_{2}=\frac{1}{Z_{2}^{\frac{1}{2}}}\left[
\begin{array}{c}
ie^{i\theta _{-}}\gamma \\
-\gamma \\
1 \\
ie^{i\theta _{-}}%
\end{array}%
\right] ;\hat{A}_{3}=\frac{1}{Z_{1}^{\frac{1}{2}}}\left[
\begin{array}{c}
ie^{i\theta _{+}} \\
-1 \\
\gamma \\
ie^{i\theta _{+}}\gamma%
\end{array}%
\right] ;\hat{A}_{4}=\frac{1}{Z_{2}^{\frac{1}{2}}}\left[
\begin{array}{c}
i\gamma \\
e^{i\theta _{-}}\gamma \\
-e^{i\theta _{-}} \\
i%
\end{array}%
\right] ,
\end{equation}%
where
\begin{eqnarray}
e^{\pm i\theta _{+}} &=&\frac{k^{+}\pm ik_{y}}{\sqrt{k^{+2}+k_{y}^{2}}}%
;e^{\pm i\theta _{-}}=\frac{k^{-}\pm ik_{y}}{\sqrt{k^{-2}+k_{y}^{2}}}, \\
Z_{1\left( 2\right) } &=&2\left[ e^{i\left( \theta \pm -\theta _{\mp
}\right) }+1\right] ,
\end{eqnarray}%
\begin{equation}
\gamma =\left\{
\begin{array}{l}
\frac{\Delta }{E+\sqrt{E^{2}-\Delta ^{2}}},E>\Delta \\
\frac{\Delta }{E+i\sqrt{\Delta ^{2}-E^{2}}},-\Delta \leq E\leq \Delta \\
\frac{\Delta }{E-\sqrt{E^{2}-\Delta ^{2}}},E<-\Delta%
\end{array}%
\right. ,
\end{equation}%
and%
\begin{equation}
v_{f}k^{\pm }=\sqrt{\left( \mu \pm \sqrt{E^{2}-\Delta ^{2}}\right)
^{2}-k_{y}^{2}}.
\end{equation}%
For the conjugate process, we have%
\begin{equation}
\tilde{H}=\left[
\begin{array}{cc}
\tilde{h}(k_{x},k_{y})+\tilde{M} & i\hat{\sigma}_{y}\Delta \\
-i\hat{\sigma}_{y}\Delta & -\tilde{h}^{\ast }(-k_{x},-k_{y})-\tilde{M}^{\ast
}%
\end{array}%
\right] ,
\end{equation}%
with $\tilde{h}(k_{x},k_{y})=v_{f}(-k_{y}\hat{\sigma}_{x}-k_{x}\hat{\sigma}%
_{y})-\mu $ and $\tilde{M}=M^{\ast }$. The eigenfunctions of $\tilde{H}$ are
$\hat{B}_{1}e^{ik^{+}x^{\prime }-ik_{y}y^{\prime }}$, $\hat{B}%
_{2}e^{-ik^{-}x^{\prime }-ik_{y}y^{\prime }}$, $\hat{B}_{3}e^{-ik^{+}x^{%
\prime }-ik_{y}y^{\prime }}$, and $\hat{B}_{4}e^{ik^{-}x^{\prime
}-ik_{y}y^{\prime }}$, with
\begin{equation}
\hat{B}_{1}=\frac{1}{Z_{2}^{\frac{1}{2}}}\left[
\begin{array}{c}
ie^{-i\theta _{+}} \\
1 \\
-\gamma \\
ie^{-i\theta _{+}}\gamma%
\end{array}%
\right] ;\hat{B}_{2}=\frac{1}{Z_{1}^{\frac{1}{2}}}\left[
\begin{array}{c}
i\gamma \\
-e^{-i\theta _{-}}\gamma \\
e^{-i\theta _{-}} \\
i%
\end{array}%
\right] ;\hat{B}_{3}=\frac{1}{Z_{2}^{\frac{1}{2}}}\left[
\begin{array}{c}
i \\
-e^{-i\theta _{+}} \\
e^{-i\theta _{+}}\gamma \\
i\gamma%
\end{array}%
\right] ;\hat{B}_{4}=\frac{1}{Z_{1}^{\frac{1}{2}}}\left[
\begin{array}{c}
ie^{-i\theta _{-}}\gamma \\
\gamma \\
-1 \\
ie^{-i\theta _{-}}%
\end{array}%
\right].
\end{equation}

\subsection{Retarded Green's function}

Now, we use McMillan's method to construct the retarded Green's function $%
G^{R}\left( x,x^{\prime },E+i\eta \right) $ which is analytic in the upper
half complex-plane. For $x>x^{\prime }$, $G_{x>x^{\prime }}^{R}\left(
x,x^{\prime },y,y^{\prime }\right) =e^{ik_{y}y-ik_{y}y^{\prime
}}G_{x>x^{\prime }}^{R}\left( x,x^{\prime }\right) $ with
\begin{equation}
G_{x>x^{\prime }}^{R}\left( x,x^{\prime }\right) =\alpha _{1}\psi _{1}\left(
x\right) \tilde{\psi}_{3}^{T}\left( x^{\prime }\right) +\alpha _{2}\psi
_{1}\left( x\right) \tilde{\psi}_{4}^{T}\left( x^{\prime }\right) +\alpha
_{3}\psi _{2}\left( x\right) \tilde{\psi}_{3}^{T}\left( x^{\prime }\right)
+\alpha _{4}\psi _{2}\left( x\right) \tilde{\psi}_{4}^{T}\left( x^{\prime
}\right) ,  \label{retarded1}
\end{equation}%
and for $x<x^{\prime }$, $G_{x<x^{\prime }}^{R}\left( x,x^{\prime
},y,y^{\prime }\right) =e^{ik_{y}y-ik_{y}y^{\prime }}G_{x<x^{\prime
}}^{R}\left( x,x^{\prime }\right) $
with
\begin{equation}
G_{x<x^{\prime }}^{R}\left( x,x^{\prime }\right) =\beta _{1}\psi _{3}\left(
x\right) \tilde{\psi}_{1}^{T}\left( x^{\prime }\right) +\beta _{2}\psi
_{4}\left( x\right) \tilde{\psi}_{1}^{T}\left( x^{\prime }\right) +\beta
_{3}\psi _{3}\left( x\right) \tilde{\psi}_{2}^{T}\left( x^{\prime }\right)
+\beta _{4}\psi _{4}\left( x\right) \tilde{\psi}_{2}^{T}\left( x^{\prime
}\right) .  \label{retarded2}
\end{equation}

\begin{figure}[!h]
\begin{center}
\includegraphics[width = 80 mm]{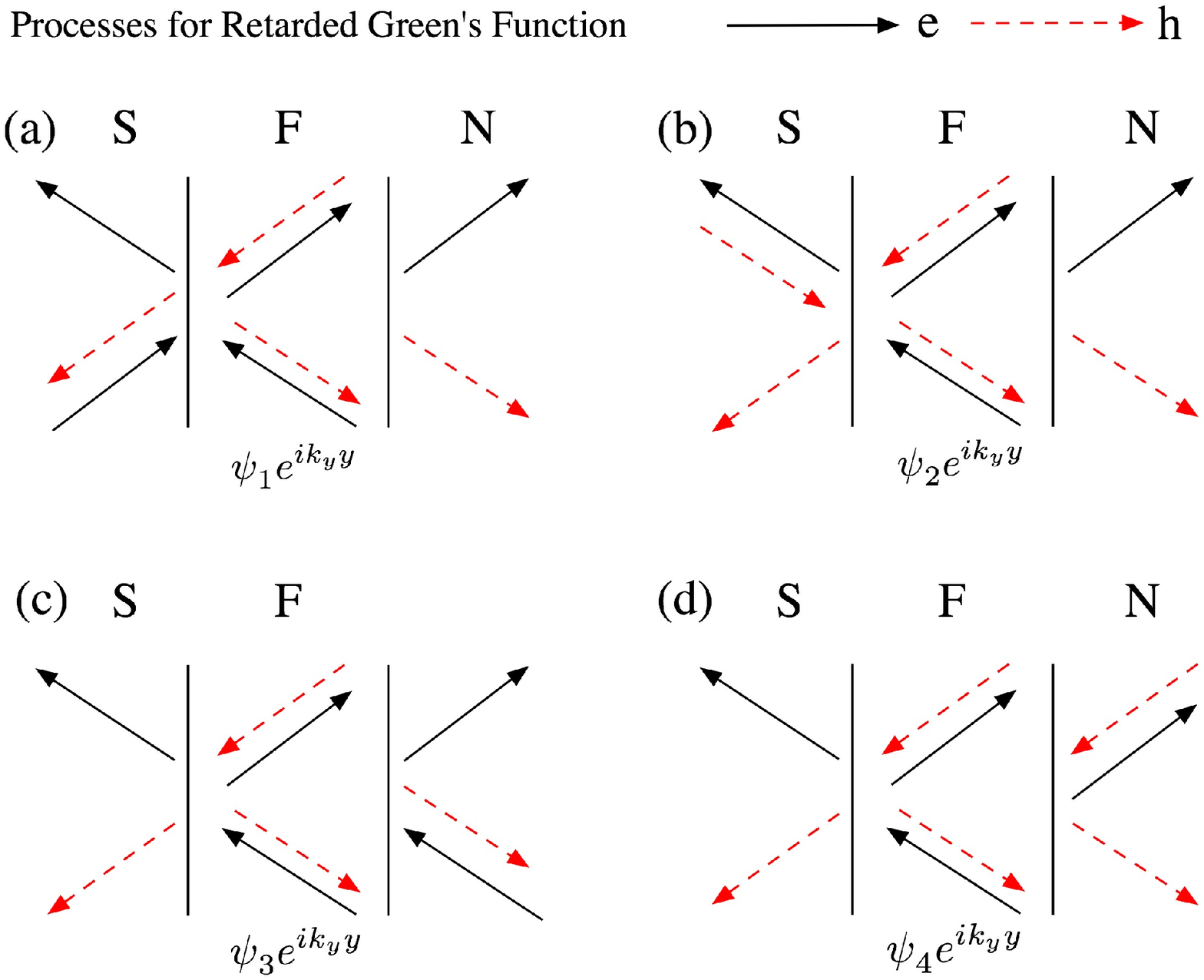}
\end{center}
\caption{Schematic illustration of four types of elementary injection
process for retarded Green's function in S/F/N junctions. $e$ and $h$ denote
electron and hole. }
\label{fig1}
\end{figure}

\begin{figure}[!h]
\begin{center}
\includegraphics[width = 80 mm]{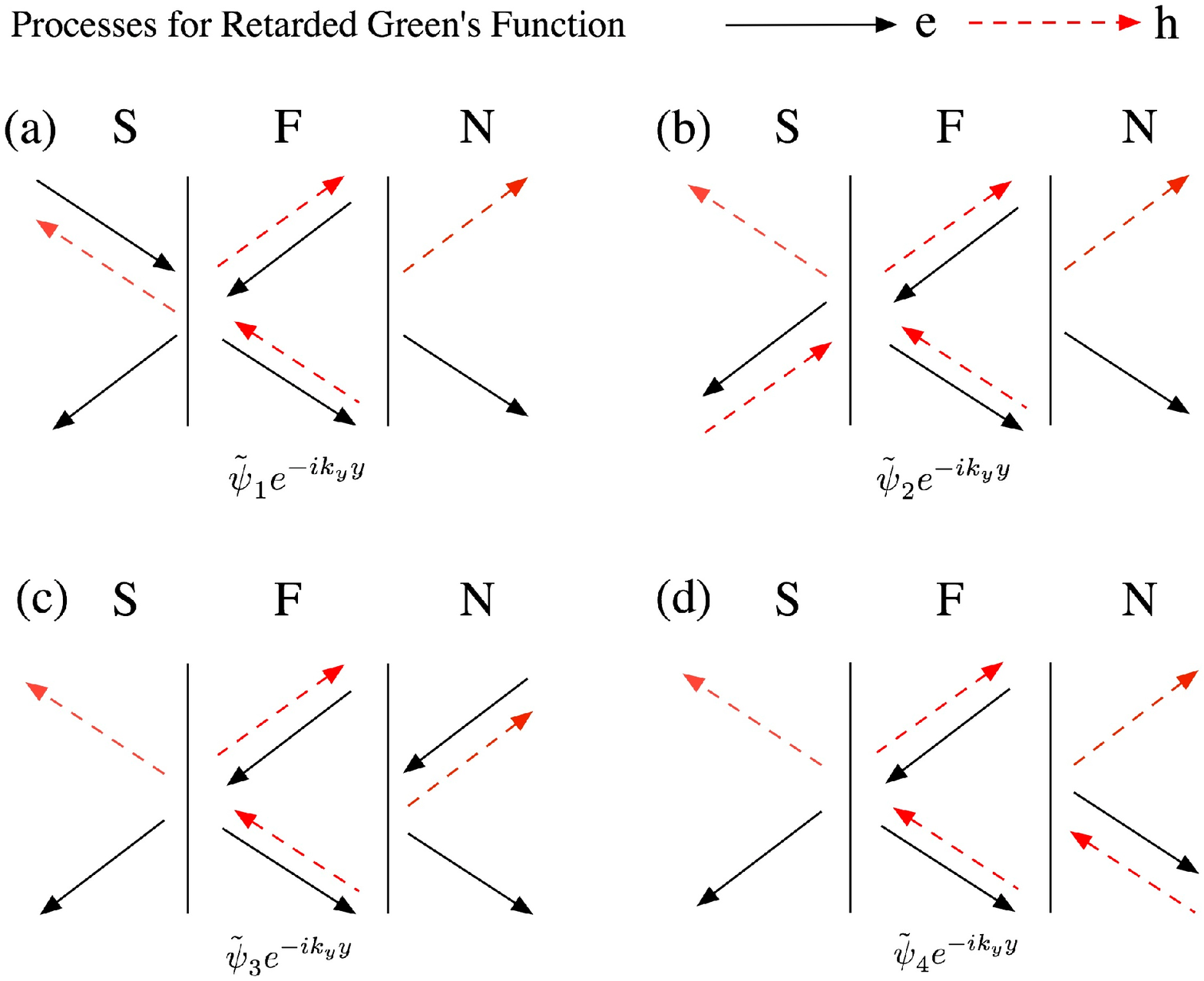}
\end{center}
\caption{Schematic illustration of four types of injection process conjugate
to those in Fig. 1. $e$ and $h$ denote electron and hole.}
\label{fig2}
\end{figure}

The wave functions
$\psi _{1}\left( x\right) $, $\psi _{2}\left( x\right) $
$\psi _{3}\left( x\right) $, and $\psi _{4}\left( x\right) $
exhibit the processes in Figs.\ref{fig1}(a), (b), (c) and (d), respectively and are given by%
\begin{eqnarray}
\psi _{1}\left( x\right) &=&\hat{A}_{1}e^{ik^{+}x}+a_{1}\hat{A}%
_{4}e^{ik^{-}x}+b_{1}\hat{A}_{3}e^{-ik^{+}x}, \\
\psi _{2}\left( x\right) &=&\hat{A}_{2}e^{-ik^{-}x}+a_{2}\hat{A}%
_{3}e^{-ik^{+}x}+b_{2}\hat{A}_{4}e^{ik^{-}x}, \\
\psi _{3}\left( x\right) &=&c_{3}\hat{A}_{3}e^{-ik^{+}x}+d_{3}\hat{A}%
_{4}e^{ik^{-}x}, \\
\psi _{4}\left( x\right) &=&c_{4}\hat{A}_{4}e^{ik^{-}x}+d_{4}\hat{A}%
_{3}e^{-ik^{+}x}.
\end{eqnarray}%
$\tilde{\psi}_{1}\left( x^{\prime }\right) $,
$\tilde{\psi}_{2}\left( x^{\prime }\right) $,
$\tilde{\psi}_{3}\left( x^{\prime }\right) $,
and $\tilde{\psi}_{4}\left( x^{\prime }\right) $
exhibit those
in Figs.\ref{fig1}(a), (b), (c) and (d), respectively.
They are given by%
\begin{eqnarray}
\tilde{\psi}_{1}\left( x^{\prime }\right) &=&\hat{B}_{1}e^{ik^{+}x^{\prime
}}+\tilde{a}_{1}\hat{B}_{4}e^{ik^{-}x^{\prime }}+\tilde{b}_{1}\hat{B}%
_{3}e^{-ik^{+}x^{\prime }}, \\
\tilde{\psi}_{2}\left( x^{\prime }\right) &=&\hat{B}_{2}e^{-ik^{-}x^{\prime
}}+\tilde{a}_{2}\hat{B}_{3}e^{-ik^{+}x^{\prime }}+\tilde{b}_{2}\hat{B}%
_{4}e^{ik^{-}x^{\prime }}, \\
\tilde{\psi}_{3}\left( x^{\prime }\right) &=&\tilde{c}_{3}\hat{B}%
_{3}e^{-ik^{+}x^{\prime }}+\tilde{d}_{3}\hat{B}_{4}e^{ik^{-}x^{\prime }}, \\
\tilde{\psi}_{4}\left( x^{\prime }\right) &=&\tilde{c}_{4}\hat{B}%
_{4}e^{ik^{-}x^{\prime }}+\tilde{d}_{4}\hat{B}_{3}e^{-ik^{+}x^{\prime }}.
\end{eqnarray}%
Substitute $\psi _{i=1\sim 4}\left( x\right) $ and $\tilde{\psi}_{i=1\sim
4}\left( x^{\prime }\right) $ into Eqs.(\ref{retarded1}) and (\ref{retarded2}%
), we can obtain
\begin{equation}
G^{R}\left( x,x^{\prime }\right) _{x>x^{\prime }}=\text{I}+\text{II}+\text{%
III,}
\end{equation}%
\begin{equation}
\text{I}=g_{1}\hat{A}_{1}\hat{B}_{3}^{T}e^{ik^{+}x-ik^{+}x^{\prime }}+g_{4}%
\hat{A}_{2}\hat{B}_{4}^{T}e^{-ik^{-}x+ik^{-}x^{\prime }},
\end{equation}%
\begin{equation}
\text{II}=g_{2}\hat{A}_{1}\hat{B}_{4}^{T}e^{ik^{+}x+ik^{-}x^{\prime }}+g_{3}%
\hat{A}_{2}\hat{B}_{3}^{T}e^{-ik^{-}x-ik^{+}x^{\prime }},
\end{equation}%
\begin{equation}
\begin{array}{c}
\text{III}=g_{1}a_{1}\hat{A}_{4}\hat{B}_{3}^{T}e^{ik^{-}x-ik^{+}x^{\prime
}}+g_{1}b_{1}\hat{A}_{3}\hat{B}_{3}^{T}e^{-ik^{+}x-ik^{+}x^{\prime }} \\
+g_{2}a_{1}\hat{A}_{4}\hat{B}_{4}^{T}e^{ik^{-}x+ik^{-}x^{\prime }}+g_{2}b_{1}%
\hat{A}_{3}\hat{B}_{4}^{T}e^{-ik^{+}x+ik^{-}x^{\prime }} \\
+g_{3}a_{2}\hat{A}_{3}\hat{B}_{3}^{T}e^{-ik^{+}x-ik^{+}x^{\prime
}}+g_{3}b_{2}\hat{A}_{4}\hat{B}_{3}^{T}e^{ik^{-}x-ik^{+}x^{\prime }} \\
+g_{4}a_{2}\hat{A}_{3}\hat{B}_{4}^{T}e^{-ik^{+}x+ik^{-}x^{\prime
}}+g_{4}b_{2}\hat{A}_{4}\hat{B}_{4}^{T}e^{ik^{-}x+ik^{-}x^{\prime }}%
\end{array}%
,
\end{equation}%
and
\begin{eqnarray}
&&G_{x<x^{\prime }}^{R}\left( x,x^{\prime }\right) =\text{I'}+\text{II'}+%
\text{III',} \\
\text{I'} &=&h_{1}\hat{A}_{3}\hat{B}_{1}^{T}e^{-ik^{+}x+ik^{+}x^{\prime
}}+h_{4}\hat{A}_{4}\hat{B}_{2}^{T}e^{ik^{-}x-ik^{-}x^{\prime }}, \\
\text{II'} &=&h_{2}\hat{A}_{4}\hat{B}_{1}^{T}e^{+ik^{-}x+ik^{+}x^{\prime
}}+h_{3}\hat{A}_{3}\hat{B}_{2}^{T}e^{-ik^{+}x-ik^{-}x^{\prime }},
\end{eqnarray}%
\begin{equation}
\begin{array}{c}
\text{III'}=h_{1}\tilde{a}_{1}\hat{A}_{3}\hat{B}%
_{4}^{T}e^{-ik^{+}x+ik^{-}x^{\prime }}+h_{1}\tilde{b}_{1}\hat{A}_{3}\hat{B}%
_{3}^{T}e^{-ik^{+}x-ik^{+}x^{\prime }} \\
+h_{2}\tilde{a}_{1}\hat{A}_{4}\hat{B}_{4}^{T}e^{ik^{-}x+ik^{-}x^{\prime
}}+h_{2}\tilde{b}_{1}\hat{A}_{4}\hat{B}_{3}^{T}e^{ik^{-}x-ik^{+}x^{\prime }}
\\
+h_{3}\tilde{a}_{2}\hat{A}_{3}\hat{B}_{3}^{T}e^{-ik^{+}x-ik^{+}x^{\prime
}}+h_{3}\tilde{b}_{2}\hat{A}_{3}\hat{B}_{4}^{T}e^{-ik^{+}x+ik^{-}x^{\prime }}
\\
+h_{4}\tilde{a}_{2}\hat{A}_{4}\hat{B}_{3}^{T}e^{ik^{-}x-ik^{+}x^{\prime
}}+h_{4}\tilde{b}_{2}\hat{A}_{4}\hat{B}_{4}^{T}e^{ik^{-}x+ik^{-}x^{\prime }}%
\end{array}%
,
\end{equation}%
where we denote%
\begin{equation}
\left\{
\begin{array}{c}
g_{1}=\alpha _{1}\tilde{c}_{3}+\alpha _{2}\tilde{d}_{4} \\
g_{2}=\alpha _{1}\tilde{d}_{3}+\alpha _{2}\tilde{c}_{4} \\
g_{3}=\alpha _{3}\tilde{c}_{3}+\alpha _{4}\tilde{d}_{4} \\
g_{4}=\alpha _{3}\tilde{d}_{3}+\alpha _{4}\tilde{c}_{4}%
\end{array}%
\right. ,\left\{
\begin{array}{c}
h_{1}=\beta _{1}c_{3}+\beta _{2}d_{4} \\
h_{2}=\beta _{1}d_{3}+\beta _{2}c_{4} \\
h_{3}=\beta _{3}c_{3}+\beta _{4}d_{4} \\
h_{4}=\beta _{3}d_{3}+\beta _{4}c_{4}%
\end{array}%
\right. .
\end{equation}%
The Green's function must satisfy the boundary conditions obtained by
integrating Eq.(\ref{pro1}) at $x=x^{\prime }$%
\begin{equation}
G\left( x+0,x\right) -G\left( x-0,x\right) =iv_{f}^{-1}\sigma _{y}\tau _{z}.
\label{boundary}
\end{equation}%
Since the right side of Eq.(\ref{boundary}) is independent of space, we can
immediately obtain II$_{x=x^{\prime }+0}$=II'$_{x=x^{\prime }-0}$, which
gives%
\begin{equation}
g_{2}\hat{A}_{1}\hat{B}_{4}^{T}=h_{2}\hat{A}_{4}\hat{B}_{1}^{T},  \label{eq1}
\end{equation}%
\begin{equation}
g_{3}\hat{A}_{2}\hat{B}_{3}^{T}=h_{3}\hat{A}_{3}\hat{B}_{3}^{T}.  \label{eq2}
\end{equation}%
By solving Eqs.(\ref{eq1}) and (\ref{eq2}), we can find
\begin{equation}
g_{2}=h_{2}=g_{3}=h_{3}=0.
\end{equation}%
Hence, the boundary condition becomes%
\begin{equation}
\text{I}_{x=x^{\prime }+0}-\text{I'}_{x=x^{\prime }-0}=iv_{f}^{-1}\sigma
_{y}\tau _{z}.
\end{equation}%
After solving this matrix equation, one can find
\begin{eqnarray}
g_{1} &=&h_{1}=\frac{i\sqrt{Z_{1}Z_{2}}}{2v_{f}\cos \theta _{+}\left(
1-\gamma ^{2}\right) }, \\
g_{4} &=&h_{4}=\frac{i\sqrt{Z_{1}Z_{2}}}{2v_{f}\cos \theta _{-}\left(
1-\gamma ^{2}\right) }.
\end{eqnarray}%
The condition that III$_{x=x^{\prime }+0}=$III'$_{x=x^{\prime }-0}$ is
guaranteed by the following detailed balanced relations,%
\begin{eqnarray}
g_{1}a_{1} &=&h_{4}\tilde{a}_{2}, \\
g_{4}a_{2} &=&h_{1}\tilde{a}_{1}, \\
g_{1}b_{1} &=&h_{1}\tilde{b}_{1}, \\
g_{4}b_{2} &=&h_{4}\tilde{b}_{2}.
\end{eqnarray}%
Thus, we have solved the retarded Green's function%
\begin{eqnarray}
G_{x>x^{\prime }}^{R}\left( x,x^{\prime }\right) &=& g_{1}[\hat{A}_{1}\hat{B}%
_{3}^{T}e^{ik^{+}(x-x^{\prime })}+a_{1}\hat{A}_{4}\hat{B}%
_{3}^{T}e^{ik^{-}x-ik^{+}x^{\prime }}+b_{1}\hat{A}_{3}\hat{B}%
_{3}^{T}e^{-ik^{+}(x+x^{\prime })}]  \qquad  \nonumber \\
&&+g_{4}[\hat{A}_{2}\hat{B}_{4}^{T}e^{-ik^{-}(x-x^{\prime })}+a_{2}\hat{A}%
_{3}\hat{B}_{4}^{T}e^{-ik^{+}x+ik^{-}x^{\prime }}+b_{2}\hat{A}_{4}\hat{B}%
_{4}^{T}e^{ik^{-}(x+x^{\prime })}],  \qquad \quad
\end{eqnarray}%
and%
\begin{eqnarray}
G_{x<x^{\prime }}^{R}\left( x,x^{\prime }\right)  &=&h_{1}[\hat{A}_{3}\hat{B}%
_{1}^{T}e^{-ik^{+}(x-x^{\prime })}+\tilde{a}_{1}\hat{A}_{3}\hat{B}%
_{4}^{T}e^{-ik^{+}x+ik^{-}x^{\prime }}+\tilde{b}_{1}\hat{A}_{3}\hat{B}%
_{3}^{T}e^{-ik^{+}(x+x^{\prime })}]   \qquad \nonumber \\
&&+h_{4}[\hat{A}_{4}\hat{B}_{2}^{T}e^{ik^{-}(x-x^{\prime })}+\tilde{a}_{2}%
\hat{A}_{4}\hat{B}_{3}^{T}e^{ik^{-}x-ik^{+}x^{\prime }}+\tilde{b}_{2}\hat{A}%
_{4}\hat{B}_{4}^{T}e^{ik^{-}(x+x^{\prime })}].  \qquad
\end{eqnarray}

\subsection{Advanced Green's function}

The advanced Green's function is analytic in the lower complex-plane $%
G^{A}\left( x,x^{\prime },E-i\delta \right) $. We can use incoming waves to
construct it.
\begin{figure}[!h]
\begin{center}
\includegraphics[width = 80 mm]{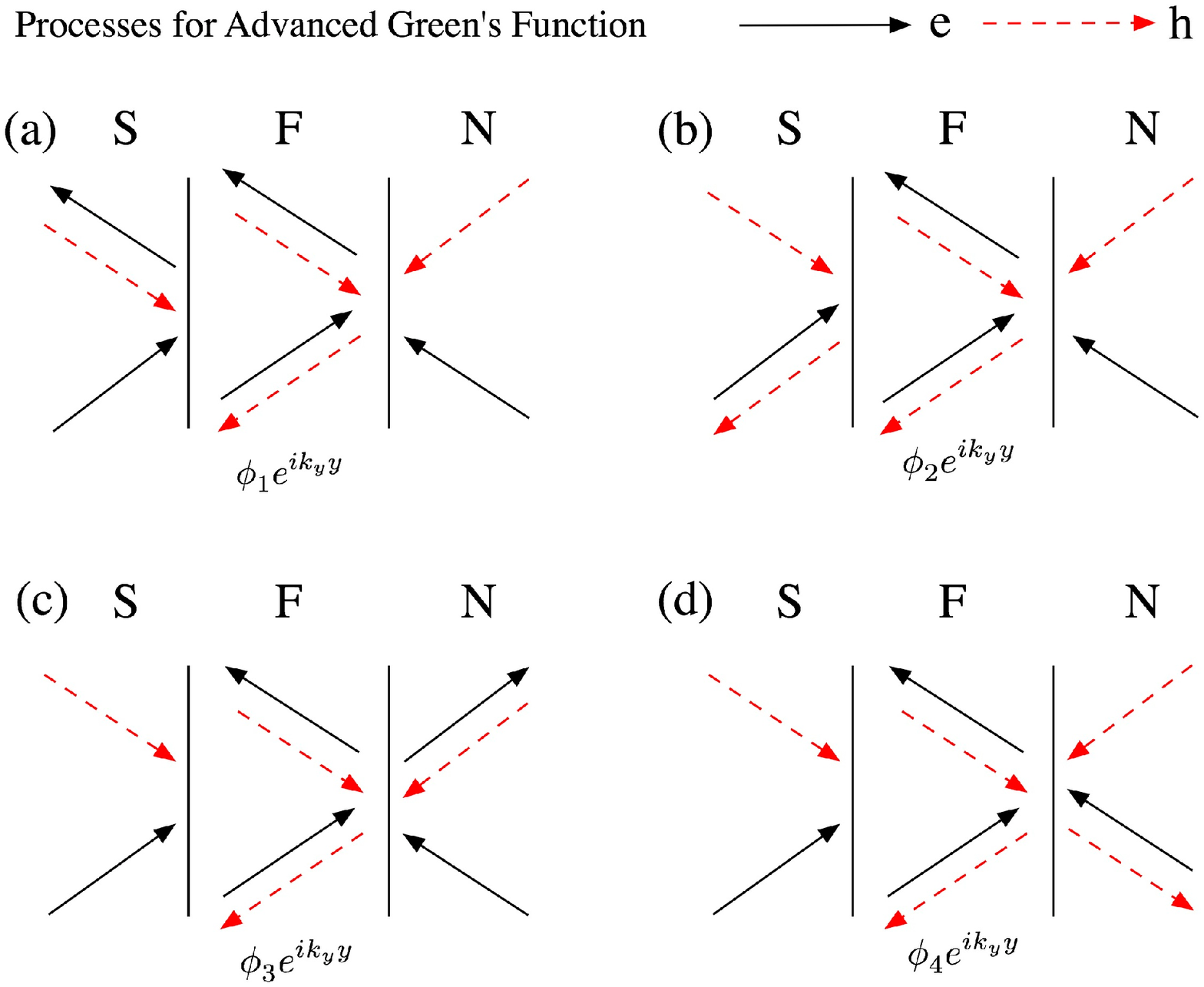}
\end{center}
\caption{Schematic illustration of four types of elementary injection
process for advanced Green's function in S/F/N junctions. $e$ and $h$ denote
electron and hole.}
\label{fig3}
\end{figure}

\begin{figure}[h]
\begin{center}
\includegraphics[width = 80 mm]{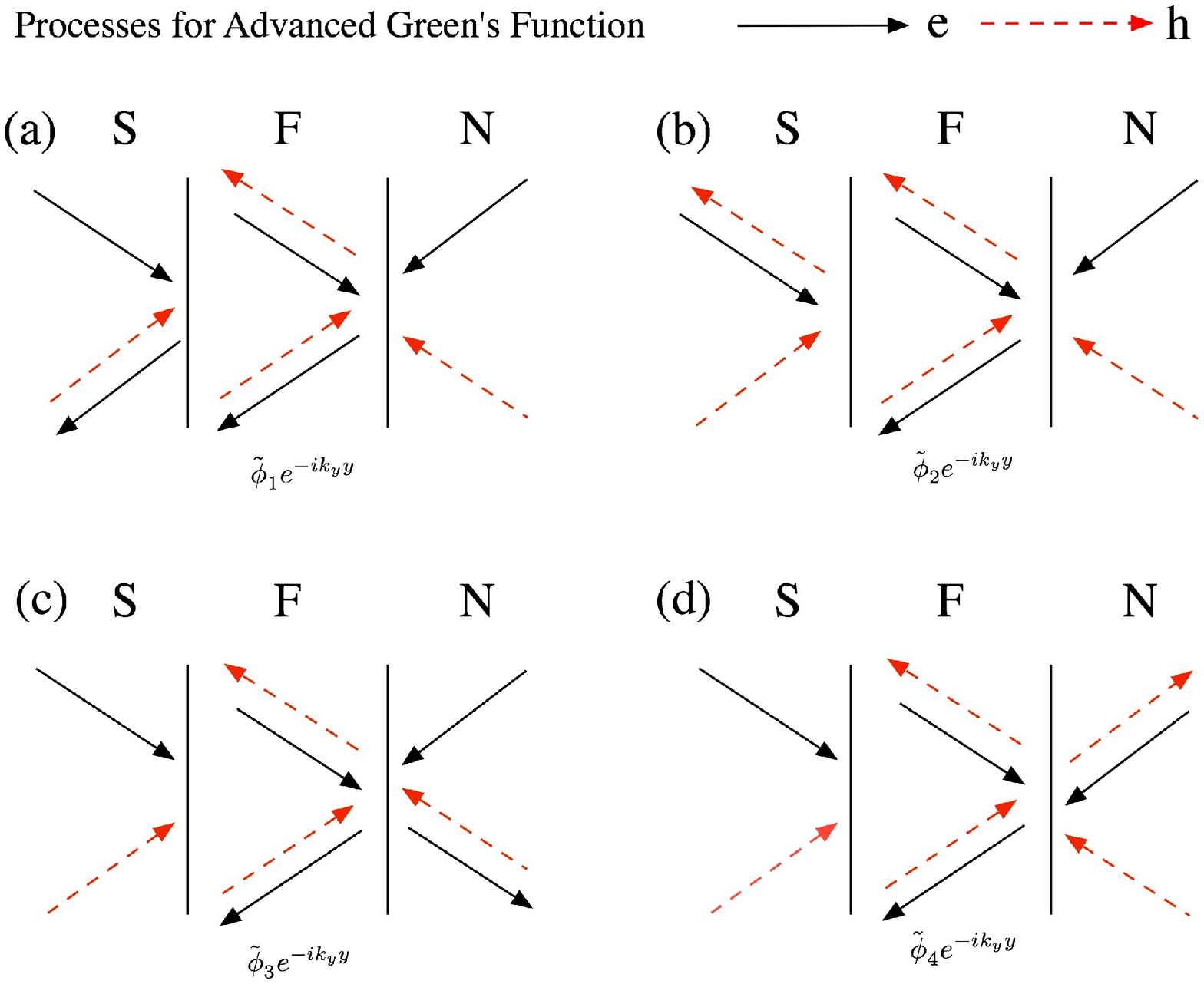}
\end{center}
\caption{Schematic illustration of four types of elementary injection
process conjugate to Fig.3 for advanced Green's function in S/F/N junctions.
$e$ and $h$ denote electron and hole.}
\label{fig4}
\end{figure}
The corresponding processes are shown in Figs.\ref{fig3} and \ref{fig4}. The
advanced Green's functions can be written as $G_{x>x^{\prime }}^{A}\left(
x,x^{\prime },y,y^{\prime }\right) =G_{x>x^{\prime }}^{A}\left( x,x^{\prime
}\right) e^{ik_{y}\left( y-y^{\prime }\right) }$ with
\begin{equation}
G_{x>x^{\prime }}^{A}\left( x,x^{\prime }\right) =\alpha _{1}^{\prime }\phi
_{1}(x)\tilde{\phi}_{3}^{T}(x^{\prime })+\alpha _{2}^{\prime }\phi _{1}(x)%
\tilde{\phi}_{4}^{T}(x^{\prime })+\alpha _{3}^{\prime }\phi _{2}(x)\tilde{%
\phi}_{3}^{T}(x^{\prime })+\alpha _{4}^{\prime }\phi _{2}(x)\tilde{\phi}%
_{4}^{T}(x^{\prime }),
\end{equation}%
and $G_{x<x^{\prime }}^{A}\left( x,x^{\prime },y,y^{\prime }\right)
=G_{x<x^{\prime }}^{A}\left( x,x^{\prime }\right) e^{ik_{y}\left(
y-y^{\prime }\right) }$ with
\begin{equation}
G_{x<x^{\prime }}^{A}\left( x,x^{\prime }\right) =\beta _{1}^{\prime }\phi
_{3}(x)\tilde{\phi}_{1}^{T}(x^{\prime })+\beta _{2}^{\prime }\phi _{4}(x)%
\tilde{\phi}_{1}^{T}(x^{\prime })+\beta _{3}^{\prime }\phi _{3}(x)\tilde{\phi%
}_{2}^{T}(x^{\prime })+\beta _{4}^{\prime }\phi _{4}(x)\tilde{\phi}%
_{2}^{T}(x^{\prime }).
\end{equation}%
In the above,
$\phi_{1}(x)$, $\phi_{2}(x)$, $\phi_{3}(x)$, and $\phi_{4}(x)$ are given by
\begin{eqnarray}
\phi _{1}\left( x\right)  &=&\hat{A}_{3}e^{-ik^{+}x}+a_{1}^{\prime }\hat{A}%
_{2}e^{-ik^{-}x}+b_{1}^{\prime }\hat{A}_{1}e^{ik^{+}x}, \\
\phi _{2}\left( x\right)  &=&\hat{A}_{4}e^{ik^{-}x}+a_{2}^{\prime }\hat{A}%
_{1}e^{ik^{+}x}+b_{2}^{\prime }\hat{A}_{2}e^{-ik^{-}x}, \\
\phi _{3}\left( x\right)  &=&c_{3}^{\prime }\hat{A}_{1}e^{ik^{+}x}+d_{3}^{%
\prime }\hat{A}_{2}e^{-ik^{-}x}, \\
\phi _{4}\left( x\right)  &=&c_{4}^{\prime }\hat{A}_{2}e^{-ik^{-}x}+d_{4}^{%
\prime }\hat{A}_{1}e^{ik^{+}x},
\end{eqnarray}%
and%
$\tilde{\phi}_{1}(x)$,
$\tilde{\phi}_{2}(x)$,
$\tilde{\phi}_{3}(x)$, and
$\tilde{\phi}_{4}(x)$ are given by
\begin{eqnarray}
\tilde{\phi}_{1}\left( x^{\prime }\right)  &=&\hat{B}_{3}e^{-ik^{+}x^{\prime
}}+\tilde{a}_{1}^{\prime }\hat{B}_{2}e^{-ik^{-}x^{\prime }}+\tilde{b}%
_{1}^{\prime }\hat{B}_{1}e^{ik^{+}x^{\prime }}, \\
\tilde{\phi}_{2}\left( x^{\prime }\right)  &=&\hat{B}_{4}e^{ik^{-}x^{\prime
}}+\tilde{a}_{2}^{\prime }\hat{B}_{1}e^{ik^{+}x^{\prime }}+\tilde{b}%
_{2}^{\prime }\hat{B}_{2}e^{-ik^{-}x^{\prime }}, \\
\tilde{\phi}_{3}\left( x^{\prime }\right)  &=&\tilde{c}_{3}^{\prime }\hat{B}%
_{1}e^{ik^{+}x^{\prime }}+\tilde{d}_{3}^{\prime }\hat{B}_{2}e^{-ik^{-}x^{%
\prime }}, \\
\tilde{\phi}_{4}\left( x^{\prime }\right)  &=&\tilde{c}_{4}^{\prime }\hat{B}%
_{2}e^{-ik^{-}x^{\prime }}+\tilde{d}_{4}^{\prime }\hat{B}_{1}e^{ik^{+}x^{%
\prime }}.
\end{eqnarray}%
Substitute $\phi_{i=1\sim 4}\left( x\right) $ and $\tilde{\phi}_{i=1\sim
4}\left( x^{\prime }\right) $  into Green's function, we have%
\begin{equation}
G_{x>x^{\prime }}^{A}\left( x,x^{\prime }\right) =\overline{\text{I}}+%
\overline{\text{II}}+\overline{\text{III}},
\end{equation}%
\begin{equation}
\overline{\text{I}}=g_{1}^{\prime }\hat{A}_{3}\hat{B}%
_{1}^{T}e^{-ik^{+}x+ik^{+}x^{\prime }}+g_{4}^{\prime }\hat{A}_{4}\hat{B}%
_{2}^{T}e^{ik^{-}x-ik^{-}x^{\prime }},
\end{equation}%
\begin{equation}
\overline{\text{II}}=g_{2}^{\prime }\hat{A}_{3}\hat{B}%
_{2}^{T}e^{-ik^{+}x-ik^{-}x^{\prime }}+g_{3}^{\prime }\hat{A}_{4}\hat{B}%
_{1}^{T}e^{ik^{-}x+ik^{+}x^{\prime }},
\end{equation}%
\begin{equation}
\begin{array}{c}
\overline{\text{III}}=g_{1}^{\prime }a_{1}^{\prime }\hat{A}_{2}\hat{B}%
_{1}^{T}e^{-ik^{-}x+ik^{+}x^{\prime }}+g_{1}^{\prime }b_{1}^{\prime }\hat{A}%
_{1}\hat{B}_{1}^{T}e^{ik^{+}x+ik^{+}x^{\prime }} \\
+g_{2}^{\prime }a_{1}^{\prime }\hat{A}_{2}\hat{B}%
_{2}^{T}e^{-ik^{-}x-ik^{-}x^{\prime }}+g_{2}^{\prime }b_{1}^{\prime }\hat{A}%
_{1}\hat{B}_{2}^{T}e^{ik^{+}x-ik^{-}x^{\prime }} \\
+g_{3}^{\prime }a_{2}^{\prime }\hat{A}_{1}\hat{B}%
_{1}^{T}e^{ik^{+}x+ik^{+}x^{\prime }}+g_{3}^{\prime }b_{2}^{\prime }\hat{A}%
_{2}\hat{B}_{1}^{T}e^{-ik^{-}x+ik^{+}x^{\prime }} \\
+g_{4}^{\prime }a_{2}^{\prime }\hat{A}_{1}\hat{B}%
_{2}^{T}e^{ik^{+}x-ik^{-}x^{\prime }}+g_{4}^{\prime }b_{2}^{\prime }\hat{A}%
_{2}\hat{B}_{2}^{T}e^{-ik^{-}x-ik^{-}x^{\prime }}%
\end{array}%
,
\end{equation}%
with%
\begin{eqnarray}
g_{1}^{\prime } &=&\alpha _{1}^{\prime }\tilde{c}_{3}^{\prime }+\alpha
_{2}^{\prime }\tilde{d}_{4}^{\prime }, \\
g_{2}^{\prime } &=&\alpha _{2}^{\prime }\tilde{c}_{4}^{\prime }+\alpha
_{1}^{\prime }\tilde{d}_{3}^{\prime }, \\
g_{3}^{\prime } &=&\alpha _{3}^{\prime }\tilde{c}_{3}^{\prime }+\alpha
_{4}^{\prime }\tilde{d}_{4}^{\prime }, \\
g_{4}^{\prime } &=&\alpha _{4}^{\prime }\tilde{c}_{4}^{\prime }+\alpha
_{3}^{\prime }\tilde{d}_{3}^{\prime },
\end{eqnarray}%
and%
\begin{equation}
G^{A}\left( x,x^{\prime }\right) _{x<x^{\prime }}=\overline{\text{I}}\text{'}%
+\overline{\text{II}}\text{'}+\overline{\text{III}}\text{',}
\end{equation}%
\begin{equation}
\overline{\text{I}}\text{'}=h_{1}^{\prime }\hat{A}_{1}\hat{B}%
_{3}^{T}e^{ik^{+}x-ik^{+}x^{\prime }}+h_{4}^{\prime }\hat{A}_{2}\hat{B}%
_{4}^{T}e^{-ik^{-}x+ik^{-}x^{\prime }},
\end{equation}%
\begin{equation}
\overline{\text{II}}\text{'}=h_{2}^{\prime }\hat{A}_{2}\hat{B}%
_{3}^{T}e^{-ik^{-}x-ik^{+}x^{\prime }}+h_{3}^{\prime }\hat{A}_{1}\hat{B}%
_{4}^{T}e^{ik^{+}x+ik^{-}x^{\prime }},
\end{equation}%
\begin{equation}
\begin{array}{c}
\overline{\text{III}}\text{'}=h_{1}^{\prime }\tilde{a}_{1}^{\prime }\hat{A}%
_{1}\hat{B}_{2}^{T}e^{ik^{+}x-ik^{-}x^{\prime }}+h_{1}^{\prime }\tilde{b}%
_{1}^{\prime }\hat{A}_{1}\hat{B}_{1}^{T}e^{ik^{+}x+ik^{+}x^{\prime }} \\
+h_{2}^{\prime }\tilde{a}_{1}^{\prime }\hat{A}_{2}\hat{B}%
_{2}^{T}e^{-ik^{-}x-ik^{-}x^{\prime }}+h_{2}^{\prime }\tilde{b}_{1}^{\prime }%
\hat{A}_{2}\hat{B}_{1}^{T}e^{-ik^{-}x+ik^{+}x^{\prime }} \\
+h_{3}^{\prime }\tilde{a}_{2}^{\prime }\hat{A}_{1}\hat{B}%
_{1}^{T}e^{ik^{+}x+ik^{+}x^{\prime }}+h_{3}^{\prime }\tilde{b}_{2}^{\prime }%
\hat{A}_{1}\hat{B}_{2}^{T}e^{ik^{+}x-ik^{-}x^{\prime }} \\
+h_{4}^{\prime }\tilde{a}_{2}^{\prime }\hat{A}_{2}\hat{B}%
_{1}^{T}e^{-ik^{-}x+ik^{+}x^{\prime }}+h_{4}^{\prime }\tilde{b}_{2}^{\prime }%
\hat{A}_{2}\hat{B}_{2}^{T}e^{-ik^{-}x-ik^{-}x^{\prime }}%
\end{array}%
,
\end{equation}%
with%
\begin{eqnarray}
h_{1}^{\prime } &=&\beta _{1}^{\prime }c_{3}^{\prime }+\beta _{2}^{\prime
}d_{4}^{\prime }, \\
h_{2}^{\prime } &=&\beta _{1}^{\prime }d_{3}^{\prime }+\beta _{2}^{\prime
}c_{4}^{\prime }, \\
h_{3}^{\prime } &=&\beta _{3}^{\prime }c_{3}^{\prime }+\beta _{4}^{\prime
}d_{4}^{\prime }, \\
h_{4}^{\prime } &=&\beta _{3}^{\prime }d_{3}^{\prime }+\beta _{4}^{\prime
}c_{4}^{\prime }.
\end{eqnarray}%
The boundary conditions of Green's function becomes,
\begin{equation}
G\left( x+0,x\right) -G\left( x-0,x\right) =iv_{f}^{-1}\sigma _{y}\tau _{z}.
\end{equation}%
Following the same procedure as in the case of calculating retarded Green's
function, we can find%
\begin{eqnarray}
g_{1}^{\prime } &=&h_{1}^{\prime }=\frac{i\sqrt{Z_{1}Z_{2}}}{2v_{f}\cos
\theta _{+}\left[ \gamma ^{2}-1\right] }, \\
g_{2}^{\prime } &=&h_{2}^{\prime }=g_{3}^{\prime }=h_{3}^{\prime }=0, \\
g_{4}^{\prime } &=&h_{4}^{\prime }=\frac{i\sqrt{Z_{1}Z_{2}}}{2v_{f}\cos
\theta _{-}\left[ \gamma ^{2}-1\right] }.
\end{eqnarray}%
The detailed balanced relations are given by
\begin{eqnarray}
g_{1}^{\prime }a_{1}^{\prime } &=&h_{4}^{\prime }\tilde{a}_{2}^{\prime }, \\
g_{4}^{\prime }a_{2}^{\prime } &=&h_{1}^{\prime }\tilde{a}_{1}^{\prime }, \\
g_{1}^{\prime }b_{1}^{\prime } &=&h_{1}^{\prime }\tilde{b}_{1}^{\prime }, \\
g_{4}^{\prime }b_{2}^{\prime } &=&h_{4}^{\prime }\tilde{b}_{2}^{\prime }.
\end{eqnarray}%
Then, the advanced Green's function is obtained by%
\begin{eqnarray}
G_{x>x^{\prime }}^{A}\left( x,x^{\prime }\right)  &=&g_{1}^{\prime }[\hat{A}%
_{3}\hat{B}_{1}^{T}e^{-ik^{+}(x-x^{\prime })}+a_{1}^{\prime }\hat{A}_{2}\hat{%
B}_{1}^{T}e^{-ik^{-}x+ik^{+}x^{\prime }}+b_{1}^{\prime }\hat{A}_{1}\hat{B}%
_{1}^{T}e^{ik^{+}(x+x^{\prime })}]  \qquad \nonumber \\
&&+g_{4}^{\prime }[\hat{A}_{4}\hat{B}_{2}^{T}e^{ik^{-}(x-x^{\prime
})}+a_{2}^{\prime }\hat{A}_{1}\hat{B}_{2}^{T}e^{ik^{+}x-ik^{-}x^{\prime
}}+b_{2}^{\prime }\hat{A}_{2}\hat{B}_{2}^{T}e^{-ik^{-}(x+x^{\prime })}],  \quad \quad \quad
\end{eqnarray}%
and%
\begin{eqnarray}
G_{x<x^{\prime }}^{A}\left( x,x^{\prime }\right)  &=&h_{1}^{\prime }[\hat{A}%
_{1}\hat{B}_{3}^{T}e^{ik^{+}(x-x^{\prime })}+\tilde{a}_{1}^{\prime }\hat{A}%
_{1}\hat{B}_{2}^{T}e^{ik^{+}x-ik^{-}x^{\prime }}+\tilde{b}_{1}^{\prime }\hat{%
A}_{1}\hat{B}_{1}^{T}e^{ik^{+}(x+x^{\prime })}]   \qquad \nonumber \\
&&+h_{4}^{\prime }[\hat{A}_{2}\hat{B}_{4}^{T}e^{-ik^{-}(x-x^{\prime })}+%
\tilde{a}_{2}^{\prime }\hat{A}_{2}\hat{B}_{1}^{T}e^{-ik^{-}x+ik^{+}x^{\prime
}}+\tilde{b}_{2}^{\prime }\hat{A}_{2}\hat{B}_{2}^{T}e^{-ik^{-}(x+x^{\prime
})}].  \qquad \quad
\end{eqnarray}

\subsection{Numerical Results}

\begin{figure}[!h]
\begin{center}
\includegraphics[width = 100 mm]{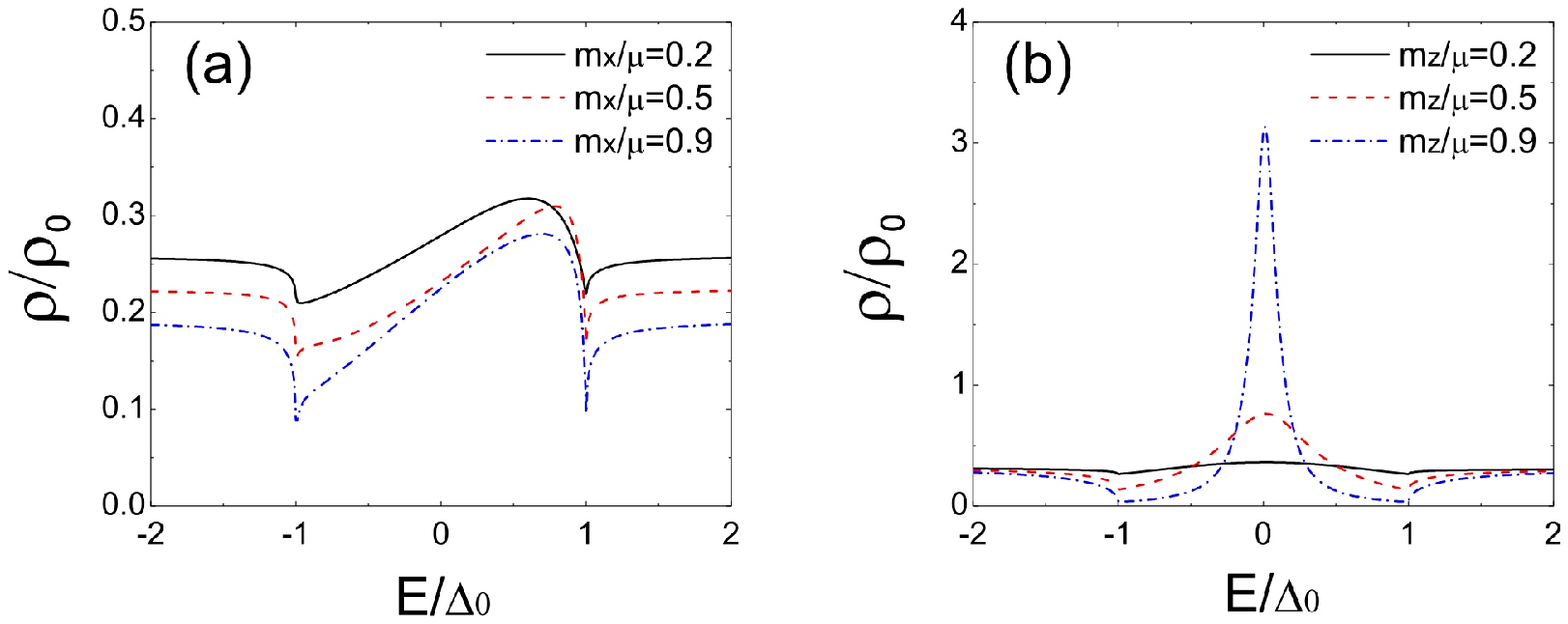}
\end{center}
\caption{Local density of states at $x=0$ in S/F/N junction.
(a) $m_{x}/\protect\mu =0.2$, $0.5,$ and $0.9$ and
(b) $m_{z}/\protect\mu =0.2$, $0.5,$ and $0.9$. $\rho_{0}$ denotes the local density of states of
bulk normal metal at Fermi energy.
Other parameters are set as $\Delta_{0}/\mu
=0.001$ and $\mu L/v_{f}=2$.}
\label{fig5}
\end{figure}

We first focus on the local density of states (LDOS) $\rho$ on the edge of
superconductor $x=0$. Here, $\rho$ is given by
\begin{equation}
\rho =-\frac{1}{\pi }\mathrm{Im}\left[ G_{11}^{R}\left( x,x\right)
+G_{22}^{R}\left( x,x\right) \right].
\end{equation}%
The obtained results are shown in Fig.\ref{fig5}. There are resonant states
inside the gap when the
magnetization is along $x$- or $z$-axis.
If resonant states are located at zero energy, they are called
Majorana Fermions. We can see when the magnetization is along $x$-axis, the
LDOS is highly asymmetric with respect to zero energy, as
shown in Fig.\ref{fig5}(a). This reveals that the in-gap resonant states
have the
characteristic similar to Shiba states, $i.e.,$ magnetic impurity state
coupled to superconductor,
in spin non-degenerate system realized
on the surface of TI \cite{Yu,Shiba,Rusinov,LuSUST}.
The LDOS becomes more prominent and
symmetric when the magnetization is along $z$-axis, as shown in Fig.\ref%
{fig5}(b). These results are consistent with previous theories of Majorana
fermion on TI and its tunneling effect \cite{Fu,Tanaka2009}.

Next, we look at the pair amplitude. We obtain the Matsubara Green's
function by analytical continuation,
\begin{eqnarray}
G^{R}\left( E+i\delta \right) &\rightarrow &G\left( i\left\vert \omega
_{n}\right\vert \right), \\
G^{A}\left( E-i\delta \right) &\rightarrow &G\left( -i\left\vert \omega
_{n}\right\vert \right).
\end{eqnarray}%
Then, we write down the Matsubara Green's function explicitly
\begin{equation}
G=\left[
\begin{array}{cccc}
G_{\uparrow \uparrow } & G_{\uparrow \downarrow } & F_{\uparrow \uparrow } &
F_{\uparrow \downarrow } \\
G_{\downarrow \uparrow } & G_{\downarrow \downarrow } & F_{\downarrow
\uparrow } & F_{\downarrow \downarrow } \\
\breve{F}_{\uparrow \uparrow } & \breve{F}_{\uparrow \downarrow } & \breve{G}%
_{\uparrow \uparrow } & \breve{G}_{\uparrow \downarrow } \\
\breve{F}_{\downarrow \uparrow } & \breve{F}_{\downarrow \downarrow } &
\breve{G}_{\downarrow \uparrow } & \breve{G}_{\downarrow \downarrow }%
\end{array}%
\right] ,
\end{equation}
The pair amplitudes at $x=x^{\prime }<0$ are given by
\begin{eqnarray}
F_{\uparrow \uparrow }^{odd} &=&\frac{\breve{F}_{\uparrow \uparrow }\left(
\omega _{n},x,x\right) -\breve{F}_{\uparrow \uparrow }\left( -\omega
_{n},x,x\right) }{2} ,\\
F_{\uparrow \downarrow }^{odd} &=&\frac{\breve{F}_{\uparrow \downarrow
}\left( \omega _{n},x,x\right) -\breve{F}_{\uparrow \downarrow }\left(
-\omega _{n},x,x\right) }{2} ,\\
F_{\downarrow \uparrow }^{odd} &=&\frac{\breve{F}_{\downarrow \uparrow
}\left( \omega _{n},x,x\right) -\breve{F}_{\downarrow \uparrow }\left(
-\omega _{n},x,x\right) }{2} ,\\
F_{\downarrow \downarrow }^{odd} &=&\frac{\breve{F}_{\downarrow \downarrow
}\left( \omega _{n},x,x\right) -\breve{F}_{\downarrow \downarrow }\left(
-\omega _{n},x,x\right) }{2}, \quad
\end{eqnarray}%
\begin{eqnarray}
F_{\uparrow \uparrow }^{even} &=&\frac{\breve{F}_{\uparrow \uparrow }\left(
\omega _{n},x,x\right) +\breve{F}_{\uparrow \uparrow }\left( -\omega
_{n},x,x\right) }{2} ,\\
F_{\uparrow \downarrow }^{even} &=&\frac{\breve{F}_{\uparrow \downarrow
}\left( \omega _{n},x,x\right) +\breve{F}_{\uparrow \downarrow }\left(
-\omega _{n},x,x\right) }{2} ,\\
F_{\downarrow \uparrow }^{even} &=&\frac{\breve{F}_{\downarrow \uparrow
}\left( \omega _{n},x,x\right) +\breve{F}_{\downarrow \uparrow }\left(
-\omega _{n},x,x\right) }{2} ,\\
F_{\downarrow \downarrow }^{even} &=&\frac{\breve{F}_{\downarrow \downarrow
}\left( \omega _{n},x,x\right) +\breve{F}_{\downarrow \downarrow }\left(
-\omega _{n},x,x\right) }{2}. \quad
\end{eqnarray}%
\begin{figure}[!h]
\begin{center}
\includegraphics[width = 100 mm]{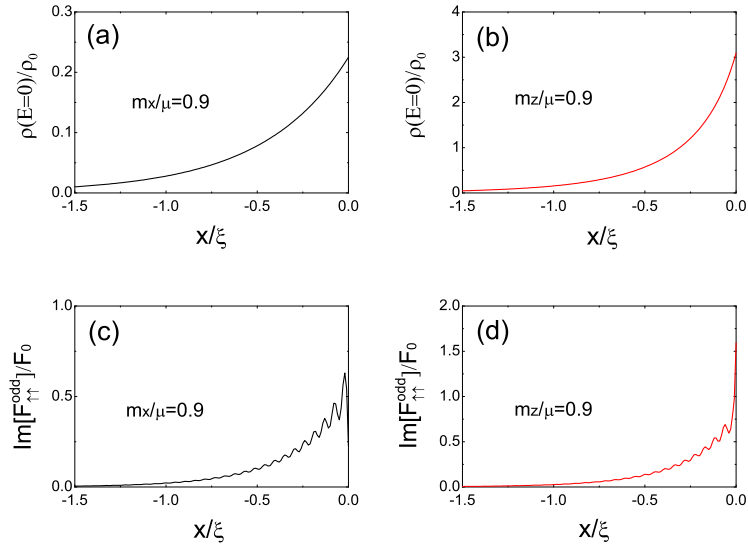}
\end{center}
\caption{Spatial dependence of zero energy local density of states,
$i.e.$, the LDOS  in the S side in S/F/N
junction for magnetization along (a) $x$-axis and (b)$z$-axis.
The spatial dependence of $F_{\uparrow \uparrow }^{odd}$
is shown in (c) and (d) corresponding to
(a) and (b), respectively. $F_{0}$ is the bulk value of $F_{\uparrow \downarrow }^{even}$
in S.
Other parameters are set as the same as in Fig. \protect\ref{fig5}.}
\label{fig6}
\end{figure}

We calculate the spatial dependence of the odd-frequency pair amplitude $%
F_{\uparrow \uparrow }^{odd}$. It is known that odd-frequency pairing is
generated by spin-rotational symmetry breaking or translational symmetry
breaking \cite{Bergeret,Tanakareview}. The magnitude of odd-frequency
pair amplitude is hugely enhanced in the presence of zero energy Andreev
bound state \cite{Tanakareview} and induces anomalous proximity effect \cite%
{Tanaka2007}. It is a remarkable feature that Majorana fermion always
accompanies odd-frequency pairing \cite{Asano,Ebisu}. Thus, it is quite
interesting to compare the relation between LDOS and
odd-frequency pairing.
Among various symmetries of odd-frequency pairings,
we focus on the odd-frequency $s$-wave pairing.
We show the results in Fig.\ref%
{fig6}. It is shown that $F_{\uparrow \uparrow }^{odd}$ is enhanced near the
S/F interface and there is a one to one
correspondence between odd-frequency pairing and Majorana fermion.
The results demonstrate the simultaneous
penetration of Majorana fermions and odd-frequency
pairing into  superconducting region.

\section{Josephson Current in S/F/S junction on TI surface}

In this section, we derive a Josephson current. We treat S/F/S junction
where the left S side is chosen as the same as that in the previous section.
The phase difference between the left and right S side is $\varphi $ so that
the pair potential in the right S is given by $\Delta e^{-i\varphi }$. The
derivation of the Josephson current is performed in the left S side. We
start from the continuity equation\cite{note}%
\begin{equation}
\frac{\partial }{\partial t}\hat{\rho}+\partial _{x}\hat{J}_{x}+\hat{S}=0,
\end{equation}%
with%
\begin{equation}
\hat{\rho}=-e\left( \hat{\Psi}_{\uparrow }^{\dag }\hat{\Psi}_{\uparrow }+%
\hat{\Psi}_{\downarrow }^{\dag }\hat{\Psi}_{\downarrow }\right) ,
\end{equation}%
\begin{equation}
\hat{J}_{x}=eiv_{f}\left[ \hat{\Psi}_{\downarrow }^{\dag }\hat{\Psi}%
_{\uparrow }-\hat{\Psi}_{\uparrow }^{\dag }\hat{\Psi}_{\downarrow }\right] ,
\end{equation}%
and
\begin{equation}
\hat{S}=ie\Delta \left[ \hat{\Psi}_{\downarrow }\hat{\Psi}_{\uparrow }-\hat{%
\Psi}_{\uparrow }^{\dag }\hat{\Psi}_{\downarrow }^{\dag }-\hat{\Psi}%
_{\uparrow }\hat{\Psi}_{\downarrow }+\hat{\Psi}_{\downarrow }^{\dag }\hat{%
\Psi}_{\uparrow }^{\dag }\right] .
\end{equation}%
Then, we make the wick rotation so that the time becomes imaginary $%
t\rightarrow i\tau $. The operators can be expressed by Matsubara Green's
function as%
\begin{eqnarray}
\left\langle {{\hat{J}_{x}}}\right\rangle  &{=}&\frac{iev_{f}}{2}{{\underset{
_{\substack{ x\rightarrow x^{\prime } \\ \tau ^{\prime }\rightarrow \tau
+0^{+}}}}{\lim }}}\left\langle {{\hat{\Psi}_{\downarrow }^{\dag }(x^{\prime
}\tau ^{\prime })\hat{\Psi}_{\uparrow }(x\tau )-\hat{\Psi}_{\uparrow }^{\dag
}(x^{\prime }\tau ^{\prime })\hat{\Psi}_{\downarrow }(x\tau )-\hat{\Psi}%
_{\uparrow }(x^{\prime }\tau ^{\prime })\hat{\Psi}_{\downarrow }^{\dag
}(x\tau )+\hat{\Psi}_{\downarrow }(x^{\prime }\tau ^{\prime })\hat{\Psi}%
_{\uparrow }^{\dag }(x\tau )}}\right\rangle   \notag \\
&=&\frac{iev_{f}}{2}{\underset{_{\substack{ x\rightarrow x^{\prime } \\ \tau
^{\prime }\rightarrow \tau +0^{+}}}}{\lim }[G_{12}(x\tau ,x^{\prime }\tau
^{\prime })-G_{21}(x\tau ,x^{\prime }\tau ^{\prime })+G_{34}(x\tau
,x^{\prime }\tau ^{\prime })-G_{43}(x\tau ,x^{\prime }\tau ^{\prime })],}
\end{eqnarray}%
\begin{eqnarray}
\left\langle \hat{S}\right\rangle  &=&ie\Delta \underset{_{\substack{ %
x\rightarrow x^{\prime } \\ \tau ^{\prime }\rightarrow \tau +0^{+}}}}{\lim }%
\left\langle \Psi _{\downarrow }\left( x^{\prime }\tau ^{\prime }\right)
\Psi _{\uparrow }\left( x\tau \right) -\Psi _{\uparrow }^{\dag }\left(
x^{\prime }\tau ^{\prime }\right) \Psi _{\downarrow }^{\dag }\left( x\tau
\right) -\Psi _{\uparrow }\left( x^{\prime }\tau ^{\prime }\right) \Psi
_{\downarrow }\left( x\tau \right) +\Psi _{\downarrow }^{\dag }\left(
x^{\prime }\tau ^{\prime }\right) \Psi _{\uparrow }^{\dag }\left( x\tau
\right) \right\rangle   \notag \\
&=&ie\Delta \underset{_{\substack{ x\rightarrow x^{\prime } \\ \tau ^{\prime
}\rightarrow \tau +0^{+}}}}{\lim }[G_{14}\left( x\tau ,x^{\prime }\tau
^{\prime }\right) -G_{23}\left( x\tau ,x^{\prime }\tau ^{\prime }\right)
+G_{32}\left( x\tau ,x^{\prime }\tau ^{\prime }\right) -G_{41}\left( x\tau
,x^{\prime }\tau ^{\prime }\right) ].
\end{eqnarray}%
For the positive frequency $\omega _{n}>0$, we can obtain the Matsubara
Green's function by analytical continuation $E+i\delta \rightarrow i\omega
_{n}$ from the retarded Green's function
\begin{eqnarray}
G_{x>x^{\prime }}^{R}\left( x,x^{\prime }\right)  &=&g_{1}\left[ \hat{A}_{1}%
\hat{B}_{3}^{T}e^{ik^{+}(x-x^{\prime })}+a_{1}\hat{A}_{4}\hat{B}%
_{3}^{T}e^{ik^{-}x-ik^{+}x^{\prime }}+b_{1}\hat{A}_{3}\hat{B}%
_{3}^{T}e^{-ik^{+}(x+x^{\prime })}\right]   \notag \\
&&+g_{4}\left[ \hat{A}_{2}\hat{B}_{4}^{T}e^{-ik^{-}(x-x^{\prime })}+a_{2}%
\hat{A}_{3}\hat{B}_{4}^{T}e^{-ik^{+}x+ik^{-}x^{\prime }}+b_{2}\hat{A}_{4}%
\hat{B}_{4}^{T}e^{ik^{-}(x+x^{\prime })}\right] .\qquad
\end{eqnarray}%
We can find
\begin{equation}
\left\langle \hat{J}_{x}\right\rangle =ev_{f}\left[ \frac{g_{1}a_{1}}{Z_{2}}%
(e^{-i\theta _{+}}+e^{i\theta _{-}})-\frac{g_{4}a_{2}}{Z_{1}}(e^{i\theta
_{+}}+e^{-i\theta _{-}})\right] \gamma e^{i\left( k^{-}-k^{+}\right) x}.
\end{equation}%
It is noted that
\begin{eqnarray}
k^{+} &=&\sqrt{\left( \mu +i\sqrt{\omega _{n}^{2}+\Delta ^{2}}\right)
^{2}-k_{y}^{2}}/v_{f}, \\
k^{-} &=&\sqrt{\left( \mu -i\sqrt{\omega _{n}^{2}+\Delta ^{2}}\right)
^{2}-k_{y}^{2}}/v_{f},
\end{eqnarray}%
we have the relations%
\begin{equation}
k^{+2}-k^{-2}=4ik_{f}\sqrt{\omega _{n}^{2}+\Delta ^{2}}/v_{f},\quad \quad
\left( \mu /v_{f}=k_{f}\right) ,
\end{equation}%
and
\begin{eqnarray}
e^{-i\theta _{+}}+e^{i\theta _{-}} &=&\frac{1}{4}\frac{k^{+}+k^{-}}{k_{f}}%
Z_{2}, \\
e^{i\theta _{+}}+e^{-i\theta _{-}} &=&\frac{1}{4}\frac{k^{+}+k^{-}}{k_{f}}%
Z_{1}.
\end{eqnarray}%
Hence, quasiparticle current becomes%
\begin{equation}
\left\langle \hat{J}_{x}\right\rangle =\frac{e\sqrt{Z_{1}Z_{2}}\left(
k^{+}+k^{-}\right) \Delta }{16\sqrt{\omega _{n}^{2}+\Delta ^{2}}}\left[
\frac{a_{1}}{k_{f}\cos \theta _{+}}-\frac{a_{2}}{k_{f}\cos \theta _{-}}%
\right] e^{i\left( k^{-}-k^{+}\right) x}.
\end{equation}%
The current which arises from the source term is given by%
\begin{eqnarray}
J_{s} &=&\int_{0}^{x}dx\left\langle \hat{S}\right\rangle =ei\Delta \left[
-g_{1}a_{1}+g_{4}a_{2}\right] \frac{\left( \gamma ^{2}-1\right) }{2}%
\int_{0}^{x}dx\left[ e^{i\left( k^{-}-k^{+}\right) x}\right]   \notag \\
&=&\frac{e\sqrt{Z_{1}Z_{2}}\left( k^{+}+k^{-}\right) \Delta }{16\sqrt{\omega
_{n}^{2}+\Delta ^{2}}}\left[ \frac{a_{1}}{k_{f}\cos \theta _{+}}-\frac{a_{2}%
}{k_{f}\cos \theta _{-}}\right] \left[ 1-e^{i\left( k^{-}-k^{+}\right) x}%
\right] .
\end{eqnarray}%
Finally, we can obtain the total Josephson current $J=\left\langle \hat{J}%
_{x}\right\rangle +J_{s}$%
\begin{equation}
J=\frac{e\sqrt{Z_{1}Z_{2}}\left( k^{+}+k^{-}\right) \Delta }{16\sqrt{\omega
_{n}^{2}+\Delta ^{2}}}\left[ \frac{a_{1}}{k_{f}\cos \theta _{+}}-\frac{a_{2}%
}{k_{f}\cos \theta _{-}}\right] ,(\omega _{n}>0).
\end{equation}%
In the same way, we obtain the Josephson current for the negative frequency
region from the analytical continuation from advanced Green's function%
\begin{equation}
J=\frac{-e\sqrt{Z_{1}Z_{2}}\left( k^{+}+k^{-}\right) \Delta }{16\sqrt{\omega
_{n}^{2}+\Delta ^{2}}}\left[ \frac{a_{1}^{\prime }}{k_{f}\cos \theta _{+}}-%
\frac{a_{2}^{\prime }}{k_{f}\cos \theta _{-}}\right] ,(\omega _{n}<0).
\end{equation}

We sum Matsubara frequencies and wave vector $k_{y}$. Then, we obtain the
Josephson current
\begin{eqnarray}
I &=&ek_{B}T\sum_{k_{y},\omega _{n}>0}\frac{\sqrt{Z_{1}Z_{2}}\left(
k^{+}+k^{-}\right) \Delta }{16\sqrt{\omega _{n}^{2}+\Delta ^{2}}}\left[
\frac{a_{1}}{k_{f}\cos \theta _{+}}-\frac{a_{2}}{k_{f}\cos \theta _{-}}%
\right]   \notag \\
&&+\sum_{k_{y},\omega _{n}<0}\frac{-\sqrt{Z_{1}Z_{2}}\left(
k^{+}+k^{-}\right) \Delta }{16\sqrt{\omega _{n}^{2}+\Delta ^{2}}}\left[
\frac{a_{1}^{\prime }}{k_{f}\cos \theta _{+}}-\frac{a_{2}^{\prime }}{%
k_{f}\cos \theta _{-}}\right] .  \label{eq6}
\end{eqnarray}%
The above equation is an extended version of Furusaki-Tsukada's formula for
Josephson current on TI surface. The physical picture is that the Josephson
current is carried via Andreev reflection which transfers (annihilates)
Cooper pair. In the quasiclassical limit, with $E$, $\Delta \ll \mu $, we
have
\begin{equation}
\begin{array}{l}
Z_{1}\approx Z_{2}\approx 2, \\
\theta _{+}\approx \theta _{-}\approx \theta , \\
k_{+}\approx k_{-}\approx k_{f}\cos \theta .%
\end{array}%
\end{equation}%
Then, the resulting Josephson current becomes%
\begin{equation}
I=\frac{ek_{B}T\Delta }{2}\sum_{k_{y}}\left[ \sum_{\omega _{n}>0}\frac{%
(a_{1}-a_{2})}{\sqrt{\omega _{n}^{2}+\Delta ^{2}}}+\sum_{\omega _{n}<0}-\frac{%
(a_{1}^{\prime }-a_{2}^{\prime })}{\sqrt{\omega _{n}^{2}+\Delta ^{2}}}\right]
.
\end{equation}%
We can derive that the contribution from negative frequencies are the same
as that from positive frequencies in the above equation. Then we have
\begin{equation}
I=ek_{B}T\Delta \sum_{k_{y},\omega _{n}>0}\frac{(a_{1}-a_{2})}{\sqrt{\omega
_{n}^{2}+\Delta ^{2}}}.
\end{equation}%

\begin{figure}[!h]
\begin{center}
\includegraphics[width = 100 mm]{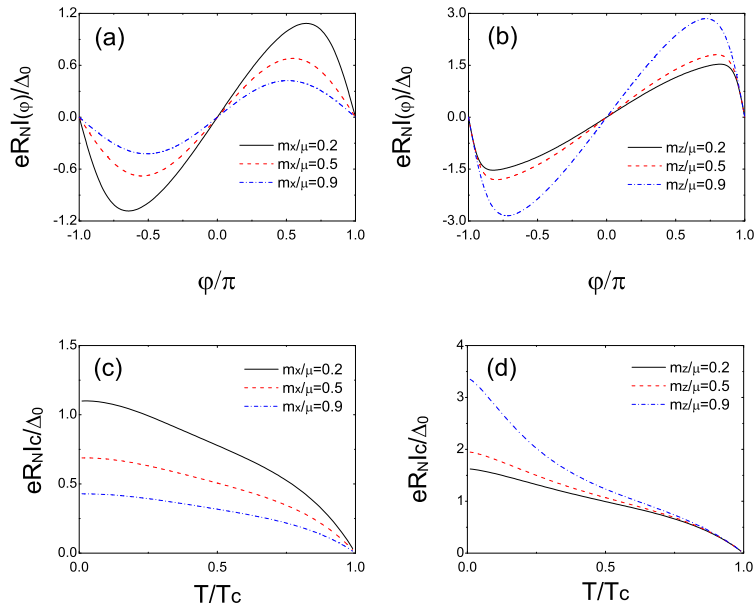}
\end{center}
\caption{Josephson current in S/F/S junction on TI surface. Current
phase relation for (a) magnetization along $x$-axis and (b) that
along $z$-axis. Temperature is set to be $T=0.1T_{c}$ where $T_{c}$ is the
critical temperature. The maximum Josephson current for magnetization along $x$- and $z$-axis is shown in (c) and (d), respectively. Other parameters
are set as the same as in Fig.\protect\ref{fig6}.}
\label{fig7}
\end{figure}

In Fig.\ref{fig7}, we present results of Josephson current in
S/F/S junctions on TI surface. It is normalized by $eR_{N}I/\Delta_{0}$ where
$R_{N}$ is the resistivity of the junction in normal state and $\Delta_{0}=\Delta (T=0)$. In panels (a)
and (b), we plot Josephson current $I(\varphi)$ as a function of the phase difference $\varphi$.
We can see that $I(\varphi)$ has a dominant first order $\sin \varphi$ coupling
in both cases independent of the direction of the
magnetization.
We also plot the maximum Josephson current $I_{c}$ with respect to $\varphi$.
Here, we assume that the temperature dependence of pair potential
$\Delta$ obeys the BCS relation:
$\Delta=\Delta _{0}\tanh (1.74\sqrt{T_{c}/T-1})$ with
$\Delta _{0}=1.76k_{B}T_{c}$ where $T_{c}$ is the critical temperature.

We plot $I_{C}$ as a function of temperature in
panels (c) and (d).
The magnitude of $eR_{N}I_{C}/\Delta_{0}$ is enhanced with the increase of
the magnetization along $z$-axis shown in panel (d),
while it becomes suppressed as the magnitude of
magnetization along $x$-axis increases, as shown in
panel (c).
In the low temperature region,
the temperature dependence of $I_{c}$ is seriously influenced by the
presence of zero-energy states.
When the magnetization is along $x$-axis, $I_{c}$ has the
Ambegaokar-Baratoff\cite{AB} type temperature dependence
for low transmissivity junction due to the absence of zero-energy states.
It is consistent with the
energy dependence of LDOS as shown in panel (a).
Although Shiba state exists at the interface,
the peak of LDOS is not located at zero energy.
Then, Shiba state does not contribute to the enhancement of Josephson current at low temperatures.
When the magnetization is along
$z$-axis, the significant
enhancement of the magnitude of LDOS at zero energy by Majorana fermion
gives rise to the
Kulik-Omelyanchuk\cite{KO} type behavior of $I_{c}$, i.e. the dash dotted line
in panel (b). Also, as compared to Fig.\ref{fig5}, we can find that the
magnitude of $I_{c}$
increases (or decreases) as the zero
energy states is enhanced (or suppressed).

\section{Discussion and Conclusion}

In this article, we have shown how to construct Green's function in superconducting hybrid systems on the
surface of topological insulator following McMillan's formalism where the
energy spectrum of electrons obeys linear dispersion. We have
obtained not only retarded
Green's function but also advanced one.
We have applied this theory for a model of superconductor/ferromagnet/normal
metal (S/F/N) junction. When the magnetization is along $x
$-axis, the local density of states is highly asymmetric around zero energy.
On the other hand, when the magnetization is along $z$-axis, we
have obtained a prominent zero-energy peak due to the generation of Majorana fermion. At
the same time, pair amplitude of odd frequency pairing is also
enhanced. We have also derived an extended Furusaki-Tsukada's formula of d.c.
Josephson current in S/F/S junctions. It is remarkable that a compact formula
of Josephson current is obtained.
When the magnetization is along $x$-axis, $I_{c}$ has the
Ambegaokar-Baratoff type temperature dependence
for low transmissivity junction.
On the other hand, when the magnetization is along
the $z$-axis, we can
obtain Kulik-Omelyanchuk type behavior of $I_{c}$
due to the presence of prominent zero-energy states by Majorana fermion.


Before closing this article, we mention about the relation between the
present theory and the quasiclassical Green's function theory. It is an
efficient way to study proximity effect based on quasiclassical Green's function theory
as far as we are focusing on the low energy excitation. Although there have
been many good review articles of quasiclassical Green's function
theory \cite{Serene,Belzig,Eschrig,Kopnin}, a theoretical procedure
to formulate quasiclassical Green's function in
superconducting junctions on TI has not been fully clarified yet\cite{Alidoust}. Only the
case in Graphene junctions has been studied up to now \cite{Takane,Beenakker1}.
Thus, it is promising to start from our obtained Green's function to establish the
quasiclassical theory of TI-based junctions and compare with the present results
in our article in future.
\enlargethispage{20pt}





\funding{This work has been supported by Topological Materials Science (TMS) (No. 15H05853)
and No. 25287085
from the Ministry of Education, Culture, Sports, Science, and
Technology, Japan (MEXT), and by the Core Research for
Evolutional Science and Technology (CREST) of the Japan
Science and Technology Corporation (JST).}

\ack{We thank P. Burset, K. Yada, A.A.Golubov and Y. Asano for valuable
discussions.}




\begin{thebibliography}{99}
\bibitem{note} We do not consider the current $\hat{J}_{y}$ since we can prove that $\left\langle
\hat{J}_{y}\right\rangle =0$ in our setup.
\end{thebibliography}
\end{document}